\documentclass[10pt,conference,a4]{IEEEtran}
\IEEEoverridecommandlockouts
\usepackage[utf8x]{inputenc}
\usepackage{amsmath,amssymb,amsfonts}
\usepackage{booktabs} 
\usepackage{cite} 
\usepackage{graphicx}
\ifCLASSOPTIONcompsoc
\usepackage[caption=false,font=normalsize,labelfont=sf,textfont=sf]{subfig}
\else
\usepackage[caption=false,font=footnotesize]{subfig}
\fi
\usepackage{url}
\usepackage[normalem]{ulem}
\usepackage{textcomp} 
\usepackage[a4paper, total={184mm,239mm}]{geometry}
\def\BibTeX{{\rm B\kern-.05em{\sc i\kern-.025em b}\kern-.08em
	T\kern-.1667em\lower.7ex\hbox{E}\kern-.125emX}}

\usepackage{scalerel}
\usepackage{tikz}
\usetikzlibrary{svg.path}

\definecolor{orcidlogocol}{HTML}{A6CE39}
\tikzset{
		orcidlogo/.pic={
				\fill[orcidlogocol] svg{M256,128c0,70.7-57.3,128-128,128C57.3,256,0,198.7,0,128C0,57.3,57.3,0,128,0C198.7,0,256,57.3,256,128z};
				\fill[white] svg{M86.3,186.2H70.9V79.1h15.4v48.4V186.2z}
				svg{M108.9,79.1h41.6c39.6,0,57,28.3,57,53.6c0,27.5-21.5,53.6-56.8,53.6h-41.8V79.1z M124.3,172.4h24.5c34.9,0,42.9-26.5,42.9-39.7c0-21.5-13.7-39.7-43.7-39.7h-23.7V172.4z}
				svg{M88.7,56.8c0,5.5-4.5,10.1-10.1,10.1c-5.6,0-10.1-4.6-10.1-10.1c0-5.6,4.5-10.1,10.1-10.1C84.2,46.7,88.7,51.3,88.7,56.8z};
			}
	}

\newcommand\orcidicon[1]{\href{https://orcid.org/#1}{\mbox{\scalerel*{
					\begin{tikzpicture}[yscale=-1,transform shape]
						\pic{orcidlogo};
						\end{tikzpicture}
				}{|}}}}

\usepackage{hyperref} 

\newcommand{\eref}[1]{(\ref{#1})}
\newcommand{\fref}[1]{Fig.~\ref{#1}}
\newcommand{\sref}[1]{Section~\ref{#1}}
\newcommand{\tref}[1]{Table~\ref{#1}}

\graphicspath{
	{images/}
}
\makeatletter
\def\@xfootnote[#1]{%
	\protected@xdef\@thefnmark{#1}%
	\@footnotemark\@footnotetext}
\makeatother

\newcommand\copyrighttext{%
	\footnotesize \textcopyright 2022 IEEE. Personal use of this material is permitted.
	Permission from IEEE must be obtained for all other uses, in any current or future
	media, including reprinting/republishing this material for advertising or promotional
	purposes, creating new collective works, for resale or redistribution to servers or
	lists, or reuse of any copyrighted component of this work in other works.
	DOI: \href{https://doi.org/10.23919/DATE54114.2022.9774757}{10.23919/DATE54114.2022.9774757}}
\newcommand\copyrightnotice{%
	\begin{tikzpicture}[remember picture,overlay]
		\node[anchor=north,xshift=0pt,yshift=-10pt] at (current page.north) {\fbox{\parbox{\dimexpr\textwidth-\fboxsep-\fboxrule\relax}{\copyrighttext}}};
	\end{tikzpicture}%
}

\begin{document}

\title{Increasing Cellular Network Energy Efficiency for Railway Corridors}

\bstctlcite{ShortCTL:BSTcontrol}

\author{
	\IEEEauthorblockN{
		Adrian Schumacher%
		\textsuperscript{*}\,\textsuperscript{\#}\orcidicon{0000-0002-6620-1871},
		Ruben Merz%
		\textsuperscript{*}\orcidicon{0000-0002-7072-5221},
		Andreas Burg%
		\textsuperscript{\#}\orcidicon{0000-0002-7270-5558}}
	\IEEEauthorblockA{
		\textsuperscript{*}
		Swisscom (Switzerland) Ltd., 3050 Bern, Switzerland;
		adrian.schumacher@swisscom.com}
	\IEEEauthorblockA{
		\textsuperscript{\#}
		Ecole polytechnique fédérale de Lausanne (EPFL), Telecommunications Circuits Laboratory, 1015 Lausanne, Switzerland}
}

\maketitle

\thispagestyle{empty}
\copyrightnotice

\begin{abstract}
  Modern trains act as Faraday cages making it challenging to provide high cellular data capacities to passengers.
A solution is the deployment of linear cells along railway tracks, forming a cellular corridor.
To provide a sufficiently high data capacity, many cell sites need to be installed at regular distances.
However, such cellular corridors with high power sites in short distance intervals are not sustainable due to the infrastructure power consumption.
To render railway connectivity more sustainable, we propose to deploy fewer high-power radio units with intermediate low-power support repeater nodes.
We show that these repeaters consume only 5\,\% of the energy of a regular cell site and help to maintain the same data capacity in the trains.
In a further step, we introduce a sleep mode for the repeater nodes that enables autonomous solar powering and even eases installation because no cables to the relays are needed.

\end{abstract}


\section{Introduction}
\label{sec:introduction}

The worldwide consumed and generated mobile data traffic is still increasing \cite{EricssonAB_EricssonMobilityReport_2021}.
Therefore, mobile network operators continuously have to upgrade their wireless infrastructure to provide sufficient data capacity and prevent data congestion.
A regular cell site consumes an average power of 3200\,W.
Because more and more cell sites are needed, people are working on optimizing the mobile network energy efficiency \cite{Falconetti_EnergyEfficiencyHeterogeneousNetworks_2012,Tombaz_ImpactDensificationEnergyEfficiency_2012,Arbi_EnergyEfficiency5GAccess_2015}.
While such optimizations are essential and needed in the long term, they are difficult because of multiple constraints and their heterogeneous structure.
An important and more structured part of the network is the infrastructure dedicated to providing cellular capacity to railways.
In fact, with the ubiquitous availability of mobile broadband, people got used to streaming media and cloud services, especially while commuting or traveling on trains.
However, regular macro networks around railway lines cannot provide the required data capacity.
The solution is a dedicated \emph{cellular corridor} providing long linear cells along railway tracks \cite{Lannoo_Radio-over-fiber-basedSolutionProvideBroadband_2007, Jamaly_DeliveringGigabitCapacitiesPassenger_2019}.
Unfortunately, providing this type of coverage to the 118,000\,km of electrified railway tracks in Europe adds up to an energy consumption of 1.24\,TWh per year.

A cellular corridor consists of masts along the railway tracks at regular inter-site distances (ISDs).
Each mast holds two antennas mounted back-to-back covering the tracks. The antennas are fed by remote radio heads (RRHs) connected to a baseband hotel via optical fiber.
Considering higher frequency bands used by the 5th generation (5G) of mobile networks and the stringent electromagnetic field (EMF) limits enforced in certain countries (e.g., Canada, Italy, Poland, Switzerland, China, Russia) \cite{Chiaraviglio_Planning5GNetworksEMF_2018}, ISDs of a few 100's of meters up to 1000\,m are necessary to provide the required data capacity.
While short ISDs are common in dense urban areas and are needed to provide sufficient cells for the capacity demand, the particular railway scenario does not need such a high cell density.
Because trains must maintain a minimum distance between each other for safety reasons, only up to one train is present in a given railway track segment at most.
Therefore, a mobile cell can be linearly stretched along the railway tracks with the help of a distributed antenna system (DAS) and multiple RRHs, reducing the number of required baseband units (BBUs).
Still, one high-power RRH can consume up to 300\,W, and with two RRHs required per site and an ISD of 500\,m, the power consumption rises to 1200\,W per kilometer of installation.
Nevertheless, ISDs of, e.g., 500\,m or more are only possible with penetration loss optimized train wagons.
Because modern train wagons significantly dampen the wireless signals \cite{Lerch_DistributedMeasurementsPenetrationLoss_2017, Jamaly_AnalysisMeasurementPenetrationLoss_2018}, which dramatically reduces data rate and capacity, onboard relays have been used to overcome the high signal attenuation.
However, active relays also consume power (650\,W for five frequency bands) and increase the required cooling energy.
Regular upgrades due to changes and extensions in the radio access network (RAN) to meet the growing data capacity demand are hard or sometimes even impossible to execute.
Therefore, structured low-emissivity (Low-E) windows, also known as frequency selective surfaces (FSS) \cite{Jamaly_AnalysisMeasurementPenetrationLoss_2018,Berisha_Smartphone-basedMeasurementsOn-boardFSS-aided_2019,Trindade_AssessmentTreatmentInfluenceMobile_2020}, have become state of the art.

\subsection{Contribution and Outline}
In this work, we propose a solution to significantly lower the high energy consumption of linear cells deployed along railway tracks.
We effectively reduce the number of complex high-power remote radio heads along the tracks without compromising cell capacity by installing energy-efficient low-power repeaters.

\begin{itemize}
	\item Based on detailed power consumption models and a calibrated model of the wireless link capacity, the system is optimized for energy efficiency while maintaining the same average data capacity as a conventional deployment.
	\item Furthermore, we exploit the network transparency of the repeaters to introduce a smart switching mechanism that allows to power these nodes with photovoltaic (PV) cells.
	\item We show that the low-power repeaters cut the average energy consumption by 50\,\% to 79\,\% compared to the conventional deployment with only high-power sites.
\end{itemize}

The rest of this article is organized as follows:
first, we list related work in \sref{sec:related}.
Then, we describe the proposed energy-efficient system architecture in \sref{sec:sysarch}.
The model and its calibration to estimate the wireless link capacity are also explained.
Moreover, we describe the used power model for cellular base stations and the power consumption of a prototype implementation for a low-power repeater node parameterized according to the same model.
Next, \sref{sec:energyautonomy} extends the power model to fully solar-powered repeater nodes and provides the corresponding energy model to show that only a few solar panels enable fully energy-autonomous operation.
Then, in \sref{sec:numeval}, the results from numerical evaluations of example deployments are presented and discussed.
Finally, \sref{sec:conclusion} concludes this paper.

\section{Related Work}
\label{sec:related}

The energy efficiency of urban heterogeneous mobile networks has been studied in \cite{Falconetti_EnergyEfficiencyHeterogeneousNetworks_2012}. Their results suggest that an existing macro network with additional low-power pico nodes improves energy efficiency compared to a conventional network with only large cell sites because users may be served from nearby pico nodes, saving energy in the macro nodes.
A similar conclusion can be drawn from \cite{Tombaz_ImpactDensificationEnergyEfficiency_2012}, where the authors found that the deployment of small cells within a macro network significantly reduces the required base station transmit power while the capacity is increased.
In contrast to the former papers, which focus on additional small cells, \cite{Arbi_EnergyEfficiency5GAccess_2015} studied if an increased macro cell sectorization could also decrease the overall energy consumption of a cellular network.
The authors compared the increased sectorization with network densification and found that additional small cells are still more energy efficient.
Those methods only improve the energy efficiency normalized by the provided capacity based on a given number of high-power macro cell sites.
However, reducing the number of high-power sites while maintaining the capacity and the specific needs of cellular railway corridors has not been studied.

The work described in \cite{Auer_CellularEnergyEfficiencyEvaluation_2011} shows how the power consumption of RAN equipment can be described accurately in a simple and handy model. Using this model, the authors also present a case study for the energy efficiency of Long Term Evolution (LTE).
An additional parameter to the mentioned model is added in \cite{Holtkamp_ParameterizedBaseStationPower_2013}, taking the cell load as a fractional value into account.
The authors in \cite{Li_PowerSavingTechniques5G_2020} also studied and applied the previously mentioned model for 5G New Radio (NR), with updated parameters.
Finally, \cite{Debaillie_FlexibleFuture-ProofPowerModel_2015} describes a more complex power model that may be better suited for 5G and 6G, but it requires many more parameters, which are hard or impossible to obtain for commercial products.

\section{System Architecture and Signal Model}
\label{sec:sysarch}

\begin{figure}
	\centering
	\includegraphics[width=\columnwidth]{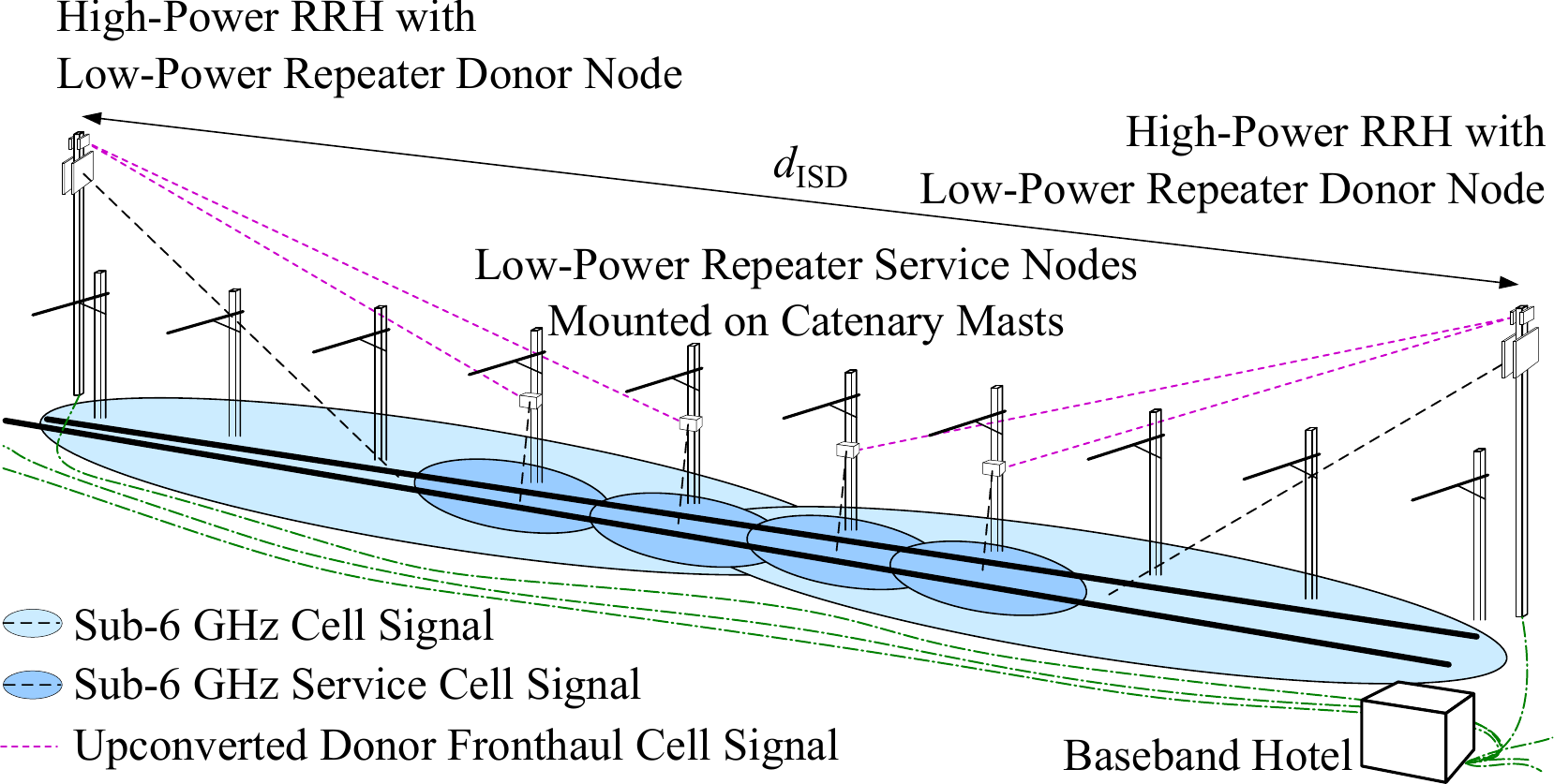}
	\caption{Railway corridor with macro sites providing the linear cell enhanced by out-of-band amplify-and-forward repeater nodes in between.}
	\label{fig:railwaycorridor}
\end{figure}

The typical deployment of a cellular corridor along a modern railway track is illustrated in \fref{fig:railwaycorridor}, with the masts for the cellular sites along the track.
Each mast is equipped with two cross-polarized pencil-beam antennas supporting two transceivers, mounted back-to-back covering the railway tracks.
Each antenna is fed by an RRH connected via optical fiber (green dash-dotted lines) to a baseband hotel.
For efficiency reasons, a single cell from a BBU is already shared by multiple RRHs along a railway track segment of several kilometers.
The data capacity is estimated from the average signal-to-noise ratio (SNR) that can be computed with the calibrated port-to-port attenuation also considering the train wagon penetration loss.
A typical deployment of this form requires high-power RRHs every 500\,m to maintain maximum data capacity provided by 5G NR inside trains.

\begin{figure}
	\centering
	\includegraphics[width=\columnwidth]{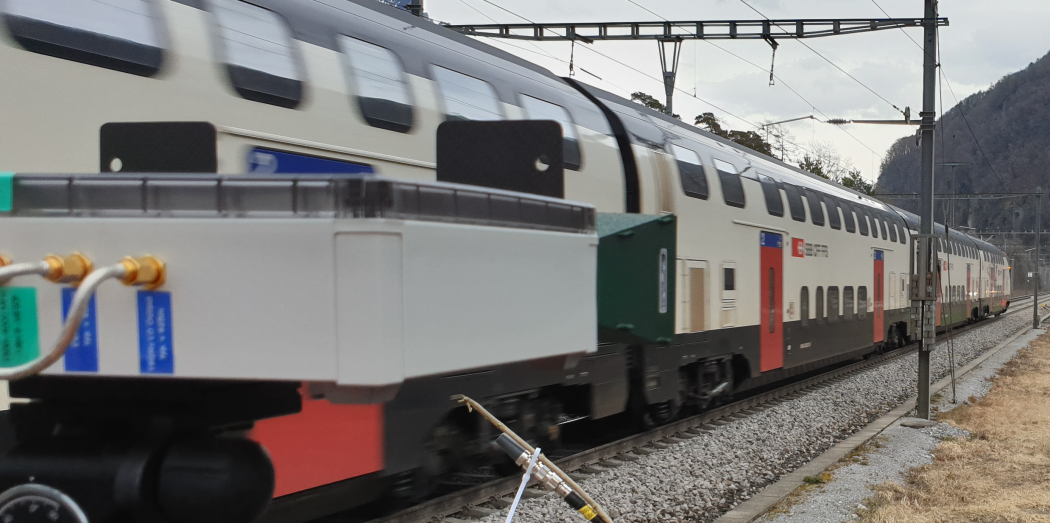}
	\caption{Low-power repeater node prototype during measurements.}
	\label{fig:hwprototype}
\end{figure}

To reduce the power consumption of the cellular corridor, we must extend the ISD of the high-power RRHs, while maintaining the system capacity.
For indoor wireless networks, it is well known that the coverage of wireless cells can be extended with signal repeaters.
However, in-band repeaters require high isolation between the antenna directed at the donor cell (e.g., macro site) and the antenna for the service cell (serving the user).
Hence, in-band repeaters are rarely considered for outdoor scenarios such as the railway track deployments considered in this paper.
Out-of-band repeaters are an attractive option to solve the isolation issue by mixing the received signal to another carrier frequency before transmitting the amplified signal.
The issue with out-of-band repeaters is that they occupy additional frequency resources, which is again prohibitive in outdoor deployments with precious licensed frequency spectrum.
The work in \cite{Schumacher_AddingIndoorCapacityFiber_2021} solves this problem by using readily available mmWave frequency bands for the donor fronthaul link that forwards the cell signal from the donor node to the service node.
In contrast to using such a mmWave bridge for mobile network capacity extension from outdoor to indoors, we propose a similar extension to reduce the number of outdoor RRHs by extending their range.
Thus, fewer high-power RRHs are needed while the same network capacity can be maintained.
A donor repeater node installed on the high-power RRH mast mixes the downlink cell signal to a higher carrier frequency for the wireless fronthaul to the service node, see \fref{fig:railwaycorridor}.
The low-power service repeater nodes mix the signal back to its original carrier frequency.
The uplink signal is treated similarly, but in the reverse direction.
For ease of deployment, the compact low-power repeater nodes can be installed on existing electric overhead wire catenary masts that are generally available every 50\,m.
We have built such a prototype and conducted several tests to validate the operation and \fref{fig:hwprototype} shows the prototype during a measurement campaign..
Further details on the prototype hardware are provided in \cite{Schumacher_AddingIndoorCapacityFiber_2021} and \cite{Schumacher_ImprovingRailwayTrackCoverage_2021}.

\subsection{Railway Track Capacity Model with Repeater Nodes}
\label{sec:capamodel}
When reducing the number of high-power macro sites while adding new low-power repeater nodes, the capacity of the system must be recalculated with a more complex model to adjust the number of repeaters to maintain the original cell capacity along the railway track.

\begin{figure}
	\centering
	\includegraphics[width=\columnwidth]{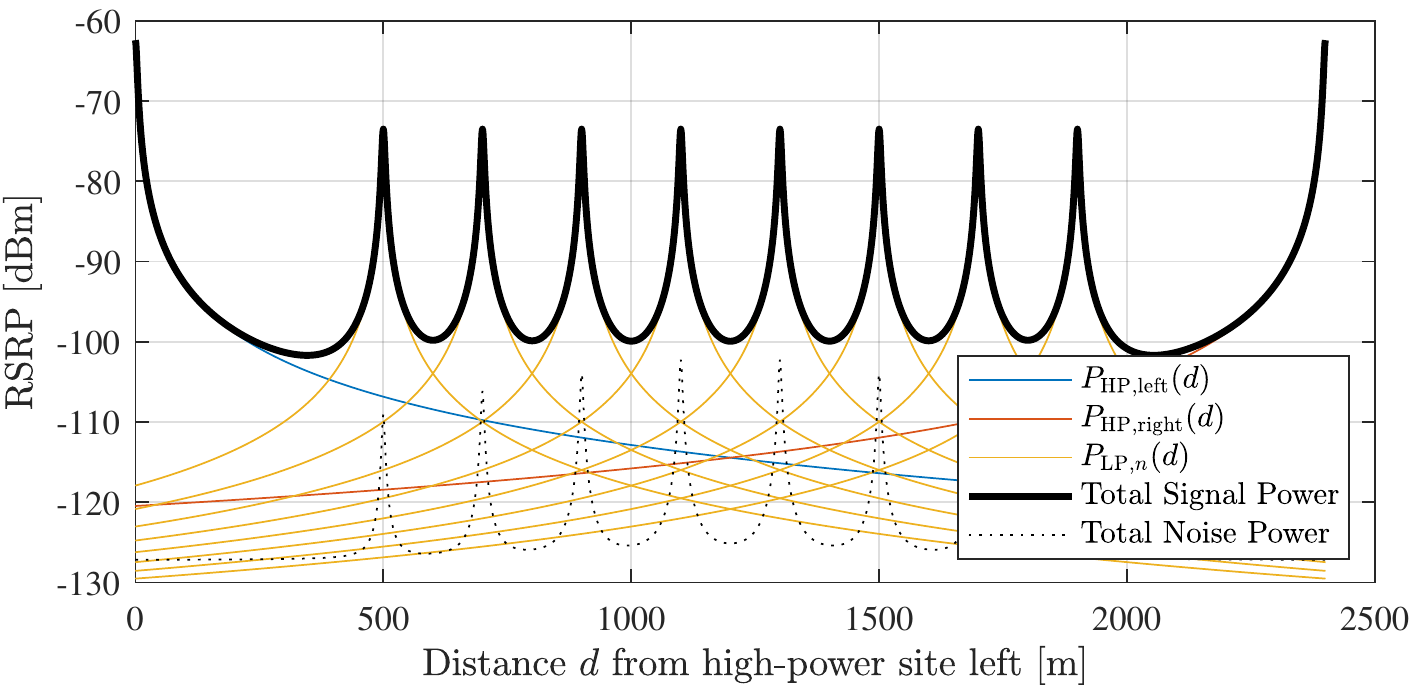}
	\caption{Signal and noise power values for $d_\textrm{ISD} = 2400$\,m and $N = 8$ low-power repeater nodes.}
	\label{fig:sysmodel_powers}
\end{figure}

Based on the high-power (HP) RRH reference signal transmit power (RSTP) $P_\textrm{HP,RSTP}$ and low-power (LP) repeater RSTP $P_\textrm{LP,RSTP}$, the reference signal received power (RSRP) along the railway track at distance $d$ can be computed.
Note that the overall signal power must be divided by the number of subcarriers to obtain the RSTP or RSRP.
Here, we consider a 5G NR carrier of 100\,MHz with 3300 subcarriers.
To obtain the RSRP for $P_\textrm{HP}(d) = P_\textrm{HP,RSTP} / L_{\textrm{HP}}(d)$ and $P_{\textrm{LP},n}(d) = P_\textrm{LP,RSTP} / L_{\textrm{LP},n}(d)$, we introduce the port-to-port attenuation $L_{\textrm{HP}}$ and $L_{\textrm{LP}}$ based on Friis with a calibration factor of $L_\textrm{HP,calib} = 33$\,dB and $L_\textrm{LP,calib} = 20$\,dB to account for antenna dependent losses into train wagons, which are in line with the measurements in \cite{Jamaly_MeasurementPathlossTrainPassengers_2021} and \cite{Schumacher_ImprovingRailwayTrackCoverage_2021}:
\begin{equation}\label{eq:fspl_fit}
	L_a (d) =
	\left( d - d_a \right)^2
	\left( \frac{4 \pi}{\lambda} \right)^2
	\cdot L_{\textrm{\{HP,LP\},calib}}.
\end{equation}
The positions of the high-power sites and the $N$ low-power repeater nodes are denoted with $d_a$, where $a \in \{ (\textrm{HP,left}), (\textrm{HP,right}), (\textrm{LP},n) | n = 1\dots N \}$, $\lambda$ represents the wavelength.
The received signal power at position $d$ from the high-power site on the left $P_\textrm{HP,left}(d) = P_\textrm{HP}(d)$ and from the high-power site on the right $P_\textrm{HP,right}(d) = P_\textrm{HP}(d_\textrm{ISD}-d)$ are illustrated in \fref{fig:sysmodel_powers} (blue and orange lines) for an example scenario.
$d_\textrm{ISD}$ denotes the inter-site distance between two high-power sites.
The received signal power from the $n$-th low-power repeater node at position $d$ is denoted by $P_{\textrm{LP},n}(d)$ and is also shown in \fref{fig:sysmodel_powers} (yellow lines).

The SNR at position $d$ along the railway track between two high-power sites is calculated according to
\begin{equation}\label{eq:ue_snr}
	\mathrm{SNR}(d) =
	\frac{ P_\textrm{HP,left}(d) + P_\textrm{HP,right}(d) +
		\sum_{n = 1}^{N} P_{\textrm{LP},n}(d) }
		{ N_\textrm{RSRP} \cdot \mathrm{NF}_\textrm{MT} +
			\sum_{n = 1}^{N} N_{\textrm{LP},n}(d) } \quad \text{.}
\end{equation}

The total noise power consists of the thermal noise floor per subcarrier $N_\textrm{RSRP} = -132$\,dBm multiplied by the noise figure of a typical mobile terminal $\mathrm{NF}_\textrm{MT} = 5$\,dB, and the received noise power from all $N$ low-power repeater nodes $N_{\textrm{LP},n}(d)$.
The noise of a low-power repeater node at distance $d$ is obtained by $N_{\textrm{LP},n}(d) = N_\textrm{RSRP} \mathrm{NF}_\textrm{LP} / L_{\textrm{LP},n}(d)$, where $\mathrm{NF}_\textrm{LP} = 8$\,dB represents the noise figure of the low-power repeater node.
The numerator (total signal power) of \eref{eq:ue_snr} and the denominator (total noise power) are both also depicted in \fref{fig:sysmodel_powers}.
If a train travels along a cellular corridor, e.g., from left to right in \fref{fig:sysmodel_powers}, a mobile terminal inside that train would see the decreasing cell signal power from the high-power site at 0\,m (blue line), which drops below -100\,dBm after around 250\,m.
In a conventional cellular corridor, the next high-power site would be located at 500\,m and starts serving the user after 250\,m with increasing power until 500\,m. With the low-power repeater nodes added instead (yellow lines), the next high-power site can be moved to, e.g., 2400\,m. Still, the signal power can be kept above $-$100\,dBm, and the mobile terminal experiences a peak for each node before the next serving high-power site dominates.

Finally, the data capacity (throughput) along the railway track can be estimated based on the above calculated SNR.
The equation in \cite[Sec.~A.2]{zz3gpp.36.942} is used, representing a calibrated model based on the Shannon bound, an attenuation factor $\alpha = 0.6$, and the maximum spectral efficiency of 5G NR $\mathrm{Thr}_\textrm{MAX} = 5.84$\,bps/Hz.

\subsection{Radio Unit Power Model}
\label{sec:powermodel}

A power consumption model for cellular infrastructure equipment has been developed in the context of the EARTH project of the European Union 7th Framework Program.
This parameterized model provides a simplified estimate of the average power consumption of cellular infrastructure equipment as a linear function of the data traffic load \cite{Wajda_EARTHDeliverable23Energy_2012}.
The power for BBU, RRH, and complete base stations, or various forms of small cells or repeaters can be approximated with the simplified equation \eref{eq:bspower}, where $P_{\textrm {in}}$ is the consumed input power.
\begin{equation}
	\label{eq:bspower}
	P_{\textrm {in}} =
	\begin{cases}
		{P_{0} +\Delta _{\textrm {p}} P_{\textrm {max}} \chi\;,} & {0 < \chi \le 1} \\
		{P_{\textrm {sleep}} \;,} & {\chi = 0}
	\end{cases}
\end{equation}
The constant $P_{0}$ represents the baseline power consumption from, e.g., power supply, oscillators, cooling.
$\Delta _{\textrm {p}}$ denotes the slope of the load-dependent power consumption, $\chi = [0,1]$ denotes the load as a fraction of the maximum possible load, and $P_{\textrm {max}}$ is the maximum radio frequency (RF) output power.
Finally, $P_{\textrm {sleep}}$ represents the constant power consumption when the equipment is in sleep mode when no data traffic is present.
The transition time between the active state and the sleep mode is assumed to be in the order of a few hundred milliseconds.

The energy consumption of cellular equipment can be calculated based on parameters from the literature
\cite{Auer_CellularEnergyEfficiencyEvaluation_2011,Wajda_EARTHDeliverable23Energy_2012,Falconetti_EnergyEfficiencyHeterogeneousNetworks_2012,Holtkamp_ParameterizedBaseStationPower_2013,Debaillie_FlexibleFuture-ProofPowerModel_2015,Li_PowerSavingTechniques5G_2020,NTTDOCOMO_BaseStationPowerModel_2011} modeled based on actual products.
With the parameters in \cite{Wajda_EARTHDeliverable23Energy_2012}, also listed in \tref{tab:pwrparams}, a high-power site consumes a power of 560\,W under full traffic load for a mast with two sectors, 336\,W under no load, and 224\,W in sleep-mode.

The power consumption for the low-power repeater node is based on the sub-components used for our prototype hardware and listed in \tref{tab:bridgepower} for common, downlink (DL), and uplink (UL) functions.
A global navigation satellite system (GNSS) disciplined oven-controlled crystal oscillator (DOCXO) is used to maintain a stable reference clock.
On the receiver (RX) side, low-noise amplifiers (LNAs) and on the transmitter (TX) side, power amplifiers (PA) are employed.
For the operation under full data traffic load, the total power consumption amounts to 28.4\,W.
If there is no data traffic, the amplifiers consume less power, but all sub-components would still be in operation, resulting in a power consumption\footnote{The power consumption of the low-power repeater node is based on a prototype, even better savings may be obtained with a fully integrated product.} of 24.3\,W.

\begin{table}
	\centering
	\caption{Low-power repeater node power consumption.}
	\label{tab:bridgepower}
	\vspace{-.5\baselineskip}
	\begin{tabular}{lrrrr}
		\toprule
		Parameter & Common [W] & DL [W] & UL [W] & Sleep [W] \\
		\midrule
		Controller & 2 & - & - & 2 \\
		GNSS DOCXO & 2.22 & - & - & 2.22 \\
		Local Oscillator & 5 & - & - & 0.5 \\
		Frequency Doubler & 0.35 & - & - & 0 \\
		RF Switches & 0.195 & - & - & 0 \\
		RX LNA & - & 0.27 & - & 0 \\
		TX PA & - & 5 & - & 0 \\
		RX LNA & - & - & 0.462 & 0 \\
		Second RX LNA & - & - & 0.335 & 0 \\
		TX PA & - & - & 5 & 0 \\
		\midrule
		Number of Path & 1 & 2 & 2 & - \\
		\midrule
		\midrule
		Total Power & \multicolumn{3}{c}{28.38} & 4.72 \\
		\bottomrule
	\end{tabular}
\end{table}

\begin{table}
	\centering
	\caption{Power model parameters for the RRH and repeater node.}
	\label{tab:pwrparams}
	\vspace{-.5\baselineskip}
	\begin{tabular}{lcccc}
		\toprule
		Node Type & $P_{\textrm {max}}$ [W] & $P_{0}$ [W] & $\Delta _{\textrm {p}}$ & $P_{\textrm {sleep}}$ [W] \\
		\midrule
		High-Power RRH & 40 & 168 & 2.8 & 112 \\
		Low-Power Repeater & 1 & 24.26 & 4.0 & 4.72 \\
		\bottomrule
	\end{tabular}
\end{table}

\section{Energy Autonomous Repeater Nodes}
\label{sec:energyautonomy}
\label{sec:sleep}

While the low-power repeater nodes already consume a fraction of a high-power RRH, the energy consumption can be further reduced by smartly switching on and off specific sub-components.
To control the whole repeater node and maintain a stable reference for the local oscillator (LO), both the low-power controller and the GNSS disciplined OCXO operate continuously.
The used LO offers a sleep mode that keeps locking onto the reference clock, but has its output drivers shut off to enable rapid on/off transitions.
All other sub-components of the repeater node are not powered in this sleep mode.
The resulting power consumption is also listed in \tref{tab:bridgepower} on the last column, and the parameters for the model in \eref{eq:bspower} are listed in \tref{tab:pwrparams}.
In this application for a cellular corridor, the sleep mode will be the primary mode of operation.
A passing train is detected using a photoelectric barrier, and the repeater node will switch to full operation during that time duration.

\subsection{Solar Power Model}
\label{sec:pvmodel}

With the intelligent sleep mode, the design and dimensioning of an off-grid PV powering system are straightforward.
The solar radiation energy captured by a PV module installed at given angles can be calculated based on equations and models documented in the literature (e.g., \cite{Calabro_AlgorithmDetermineOptimumTilt_2013}).
Several models exist for the solar radiation available at a given geographical location, either based on local measurements or estimation methods.
For our study, we use PVGIS \cite{Huld_NewSolarRadiationDatabase_2012} belonging to the latter.
The accuracy of PVGIS has been validated in several papers (e.g., \cite{AnaMariaGracia_PerformanceComparisonDifferentModels_2013,Kallioglu_EmpiricalCalculationOptimalTilt_2020a,GonzalezPena_PhotovoltaicPredictionSoftwareEvaluation_2021}) and delivers robust results.
PVGIS is a free online tool (\url{https://ec.europa.eu/jrc/en/pvgis}) providing several calculations and databases. It can also be used to calculate the performance of an off-grid PV system as it would be used for a repeater node.
Multiple statistics are calculated based on a solar radiation database, a geographical location, the peak power of the PV module and its azimuth and tilt angle, the battery capacity, and the expected daily energy consumption.

\subsection{Geographical Setting}
\label{sec:geoparams}

In our study, we consider high-speed railway corridors and evaluate four locations as examples: Madrid (Spain), Lyon (France), Vienna (Austria), Berlin (Germany).
Standard PV modules with around 0.6\,m width and 1.4\,m height providing 180\,$\textrm{W}_\textrm{p}$ are considered, which can be vertically installed on the existing catenary masts. Three modules easily fit vertically on regular masts with a height of 7\,m to 9\,m.
The following parameters were used for PVGIS:
\begin{itemize}
	\item Solar radiation database: PVGIS-COSMO
	\item Installed peak PV power: 540\,$\textrm{W}_\textrm{p}$ 
	\item Battery capacity: 720\,Wh 
	\item Discharge cutoff limit: 40\,\%
	\item Tilt/slope angle: 90\,\textdegree
	\item Azimuth angle: 0\,\textdegree
\end{itemize}

\section{Numerical Results and Discussion}
\label{sec:numeval}

Before we can calculate the energy advantages of our solution, we first need to evaluate how much the ISD of high-power masts can be extended with how many low-power repeater nodes.
Based on the path loss and capacity models in \sref{sec:capamodel}, the throughput can be calculated for every scenario (ISD in 50\,m steps, number of low-power repeater nodes \{0,$\ldots$,10\}).
For each number of nodes, the maximum ISD is registered with which the throughput still matches the peak throughput of 5G NR at an SNR $>$ 29\,dB.
Each high-power RRH and antenna is assumed to transmit with 2500\,W equivalent isotropic radiated power (EIRP) (64\,dBm), while the low-power repeater nodes transmit with a maximum of 10\,W EIRP (40\,dBm).
The resulting maximum ISDs for one to ten nodes are: \{1250, 1450, 1600, 1800, 1950, 2100, 2250, 2400, 2500, 2650\}\,m.

\subsection{Mains Powered Operation}

\begin{figure}
	\centering
	\includegraphics[width=\columnwidth]{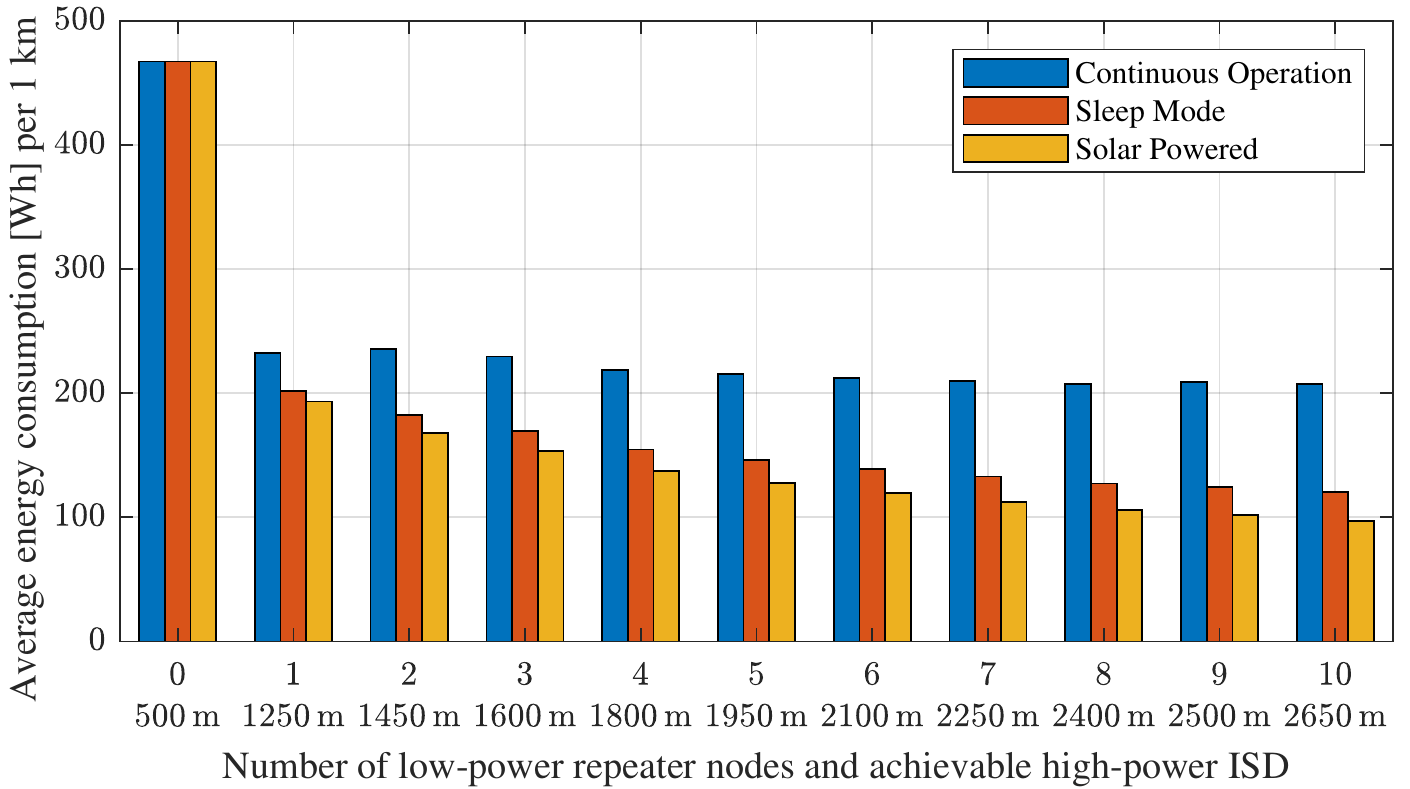}
	\vspace{-1\baselineskip}
	\caption{Average energy consumption normalized to 1\,km for the conventional cellular corridor on the very left and 1 to 10 low-power repeater nodes placed in between high-power masts (always using energy-saving techniques) to extend their ISD while maintaining the peak throughput for users.}
	\label{fig:avgenergy}
\end{figure}

We assume that high-power RRHs deployed for the specific use case of a cellular corridor along railway tracks employ power-saving functions when there is no data traffic.
Specifically, the RRH is only operating with full data traffic load for the time a train needs to pass the coverage section of the mast.
The resulting average energy consumption is caluculated according to \sref{sec:powermodel} and shown in \fref{fig:avgenergy} for different cases.
In the first case, we keep the low-power repeater nodes operating continuously, while only the RRH is switched to sleep mode if there is no traffic.
Given the parameters in \tref{tab:corridorparams}, the RRHs are in full load operation on a 24-hour average for 2.85\,\% of the time at a 500\,m high-power RRH ISD and 9.66\,\% at a 2650\,m high-power RRH ISD.
The use of at least three low-power repeater nodes extends the high-power ISD to a minimum of 1600\,m which reduces the average energy consumption per hour and kilometer to below 50\,\% of what a conventional deployment with only high-power RRHs would use (left/blue columns in \fref{fig:avgenergy}).
Note that an additional low-power repeater node as donor node is considered for one service node and two low-power donor nodes are considered for two or more service nodes (see \fref{fig:railwaycorridor}).

\begin{table}
	\centering
	\caption{Parameters for average energy consumption calculations.}
	\label{tab:corridorparams}
	\vspace{-.5\baselineskip}
	\begin{tabular}{lr}
		\toprule
		Parameter & Value \\
		\midrule
		Number of trains/h & 8 \\
		Operation under full load per train & 16\,s - 55\,s \\
		Hours per night without passenger railway traffic & 5\,h \\
		Length of a train & 400\,m \\
		LP repeater node spacing & 200\,m \\
		Velocity of a train & 200\,km/h \\
		Power for HP RRH under full load & 560\,W \\
		Power for HP RRH in sleep mode & 224\,W \\
		Power for LP repeater node under full load & 28.4\,W \\
		Power for LP repeater node no load & 24.3\,W \\
		Power for LP repeater node in sleep mode & 4.7\,W \\
		\bottomrule
	\end{tabular}
\end{table}

Using the method described earlier to reduce further the low-power repeater node energy consumption, more significant energy savings can be achieved.
One low-power repeater node then only consumes an average power of 5.17\,W (124.1\,Wh per day). According to \fref{fig:avgenergy} (middle/orange columns), a single repeater node allows extending the high-power ISD to 1250\,m yielding energy savings of 57\,\%. With ten low-power repeater nodes, an ISD of 2650\,m is achieved, resulting in 74\,\% of energy reduction.

\subsection{Energy Autonomous Repeater Operation}

If the average power consumption is small enough (roughly a single-digit Watt number), off-grid powering by a solar PV system is feasible.
As described in \sref{sec:pvmodel}, up to three standard PV modules can be mounted vertically on the railway catenary masts.
The captured solar power is stored in a battery that supplies power to the low-power repeater nodes.
According to \tref{tab:corridorparams}, the hourly energy consumption profile for PVGIS has been set to 5\,h per night continuously in sleep mode while the low-power repeater nodes operate in a mix of sleep mode and full load for the remaining 19\,h per day.
With PVGIS and the parameters listed in \sref{sec:geoparams} for high-speed railway corridors in the four exemplary regions, the performance of the autonomous PV system is analyzed.

\begin{table}
	\centering
	\caption{PVGIS results at the four exemplary regions for one year.}
	\label{tab:solarresult}
	\vspace{-.5\baselineskip}
	\begin{tabular}{@{}lrrrr@{}}
		\toprule
		Parameter & Madrid & Lyon & Vienna & Berlin \\
		\midrule
Required peak PV power [Wp] & 540 & 540 & 540 & 600 \\
Required battery capacity [Wh] & 720 & 720 & 1440 & 1440 \\
Days with full battery [\%] & 98.13 & 95.15 & 93.73 & 88.0 \\
		\bottomrule
	\end{tabular}
\end{table}

With the standard system dimensions listed in \sref{sec:geoparams}, the monthly solar radiation prediction data showed that the PV system in Vienna and Berlin might lead to an insufficient energy capacity in winter.
Therefore, the requirement of zero-days down-time dictates an adaptation. By doubling the battery capacity in Vienna and Berlin, and slightly larger PV modules for Berlin, continuous operation even in winter can be achieved.
The results are given in \tref{tab:solarresult}.

If all low-power repeater nodes are solar-powered, the overall energy consumption is further reduced because no repeater node consumes mains power, and only the high-power RRHs at extended ISDs remain mains powered.
With just one intermediate low-power repeater node, 59\,\% less energy is consumed, and with ten low-power repeater nodes between two high-power sites, 79\,\% less energy is consumed compared to the conventional deployment with only high-power RRHs, (right/yellow column in \fref{fig:avgenergy}).

\section{Conclusion}
\label{sec:conclusion}

Linear cellular corridors with high-power cell sites at regular distances are a solution to provide high data capacity along railway tracks.
However, to achieve a high data capacity onboard trains, short inter-site distances of a few hundred meters to one kilometer are needed.
Because a high-power remote radio head (RRH) consumes several hundred Watts in power, we aim to reduce the number of RRHs while maintaining the same data capacity.
We propose to install low-power out-of-band repeater nodes in between high-power sites.
The local increase in cell signal power allows extending the high-power site distances, thus requiring fewer sites and lowering the overall energy consumption.
We have built a low-power repeater node to validate the feasibility and calibrate the required various models.
We show that the overall energy consumption for a cellular corridor can be significantly lowered when deploying low-power repeater nodes between the high-power cell sites.
Further, reducing the average power consumption of a low-power repeater node allows for energy-autonomous repeater operation.
With only the high-power RRHs grid-powered but with longer inter-site distances, an energy saving of 79\,\% can be achieved.

\bibliographystyle{IEEEtran}
\bibliography{ieeebibrefstyle,references,specs3gpp,IEEEabrv}

\end{document}